\documentclass{webofc}
\usepackage[varg]{txfonts}   
\usepackage{caption}[=v3.5]
\usepackage{topcapt}
\usepackage{array}
\usepackage{xspace}
\usepackage{url}
\usepackage{breakurl}
\usepackage[breaklinks]{hyperref}

\providecommand{\cmsTable}[1]{\resizebox{\textwidth}{!}{#1}}
\newcolumntype{A}{r@{\hspace{0em}}l}
\newcolumntype{B}{A@{$\,\pm\,$}A}

\DeclareMathOperator{\bnode}{b}
\DeclareMathOperator{\bbnode}{bb}
\DeclareMathOperator{\lepbnode}{lepb}
\DeclareMathOperator{\cnode}{c}
\DeclareMathOperator{\udsnode}{uds}
\DeclareMathOperator{\gnode}{g}

\DeclareMathOperator{\B}{B}
\DeclareMathOperator{\CvB}{CvB}
\DeclareMathOperator{\CvL}{CvL}
\DeclareMathOperator{\QG}{QG}

\DeclareMathOperator{\mmd}{MMD}
\DeclareMathOperator{\mse}{MSE}

\DeclareMathOperator{\huber}{Huber}

\newcommand{\PYTHIAEIGHT} {{\textsc{pythia8}}\xspace}
\newcommand{\GEANT} {{\textsc{Geant4}}\xspace}
\newcommand{\pt}{\ensuremath{p_{\mathrm{T}}}\xspace}
\newcommand{\kt}{\ensuremath{k_{\mathrm{T}}}\xspace}
\newcommand{\GeV}{\ensuremath{\,\text{Ge\hspace{-.08em}V}}\xspace}
\newcommand{\xfast}{\ensuremath{\vec{x}^{\text{\,Fast}}}\xspace}
\newcommand{\xfull}{\ensuremath{\vec{x}^{\text{\,Full}}}\xspace}
\newcommand{\xrefined}{\ensuremath{\vec{x}^{\text{\,Refined}}}\xspace}
\newcommand{\ttbar}{\ensuremath{\text{t}\overline{\text{t}}}\xspace}

\begin{document}

\title{Refining fast simulation using machine learning}

\author{
    \firstname{Samuel} \lastname{Bein}\inst{1} \and
    \firstname{Patrick} \lastname{Connor}\inst{1,2} \and
    \firstname{Kevin} \lastname{Pedro}\inst{3} \and
    \firstname{Peter} \lastname{Schleper}\inst{1} \and
    \firstname{Moritz} \lastname{Wolf}\inst{1}\fnsep\thanks{\email{moritz.wolf@cern.ch}} \\ (on behalf of the CMS Collaboration)
}

\institute{University of Hamburg, Institut für Experimentalphysik, Germany \and
Center for Data and Computing in Natural Sciences, Hamburg, Germany \and
Fermi National Accelerator Laboratory, Batavia, IL, USA
}

\abstract{%
At the CMS experiment, a growing reliance on the fast Monte Carlo application (FastSim) will accompany the high luminosity and detector granularity expected in Phase 2. The FastSim chain is roughly 10 times faster than the application based on the \GEANT detector simulation and full reconstruction referred to as FullSim. However, this advantage comes at the price of decreased accuracy in some of the final analysis observables. In this contribution, a machine learning-based technique to refine those observables is presented. We employ a regression neural network trained with a sophisticated combination of multiple loss functions to provide post-hoc corrections to samples produced by the FastSim chain. The results show considerably improved agreement with the FullSim output and an improvement in correlations among output observables and external parameters. This technique is a promising replacement for existing correction factors, providing higher accuracy and thus contributing to the wider usage of FastSim.
}

\maketitle

\section{Introduction}
\label{sec:introduction}
Simulating particle collisions, the subsequent detector response, and the reconstruction of the final state are crucial for modern high energy physics. For the purpose of simulating events in the CMS detector~\cite{Chatrchyan:2008zzk}, the collaboration largely relies on a simulation chain based on \GEANT~\cite{Agostinelli:2002hh,Allison:2016lfl}, referred to as FullSim, which gives an accurate representation of the truth~\cite{Lange:2015sba,Hildreth:2017vpw}. However, this requires a considerable amount of computing power. Therefore, another simulation chain has been established, which uses approximations to speed up the process by roughly a factor of 10~\cite{Abdullin:2011zz, Giammanco:2014bza, Sekmen:2016iql}. This application, known as FastSim, provides output with the same format and structure as FullSim. Looking towards the future with higher LHC luminosity and increased CMS detector granularity~\cite{CMS-TDR-019}, FastSim will only gain in importance as the collaboration strives to keep the computing needs within budget in Phase 2~\cite{Pedro:2020kbk,Software:2815292}.

The output of the FastSim chain is generally in good agreement with the FullSim chain, but discrepancies on the order of up to 20\% are observed in some analysis observables. Traditionally, differences between simulation samples (or differences between simulation and data) are treated with dedicated correction factors or weights that are derived either by physics object groups or by individual users carrying out analyses. These corrections or weights are applied to events or physics objects to correct certain biases, for example in the transverse momentum or selection efficiency of jets, photons, or leptons. One approach that goes beyond these traditional corrections in terms of sophistication and accuracy is the application of weights that are derived using machine learning-based methods, such as the DCTR (deep neural networks using classification for tuning and reweighting) approach~\cite{PhysRevD.101.091901}. While the accuracy of simulated variables, as well as correlations among the variables, is improved compared to the unweighted sample, the use of weights reduces the statistical power, undermining the advantage of fast simulation applications. In contrast to reweighting, the aim of our method is to change the values of the sample to reach better agreement with the target sample, without the need for weights. Such a \emph{refinement} has been studied and tested in air shower images using a Wasserstein GAN~\cite{Erdmann:2018kuh}, demonstrating the possibility of improving the accuracy of fast particle reconstruction. We have developed a distinct method that operates on the level of final analysis observables and is realized by employing a simple regression neural network. This approach may be considered a natural replacement for the traditional correction factors currently applied to FastSim.

\section{Data sample}
\label{sec:data sample}
For the purpose of this study, the pair production of supersymmetric partners of gluons, namely, gluinos, is simulated. Each gluino subsequently decays to a top quark pair and a neutralino (supersymmetric simplified model T1tttt~\cite{LHCNewPhysicsWorkingGroup:2011mji}). Events are simulated with both the FullSim and the FastSim chain up to and including the step producing NanoAOD output~\cite{Peruzzi_2020}, which consists of high-level observables for various physics objects. Both cases share the same event generation using \PYTHIAEIGHT, referred to as the GEN step. Therefore, after the detector response and the reconstruction have been simulated, a matching can be established between a true generator-level jet and its reconstructed counterparts in the FullSim and FastSim samples; all jets are clustered using the anti-\kt algorithm~\cite{Cacciari:2008gp,Cacciari:2011ma} with a distance parameter of 0.4. The matching is achieved with a distance matching criterion of $\Delta R = \sqrt{\Delta \eta^2 + \Delta \phi^2} < 0.2$. In this way, jet triplets are constructed in the form (GEN, FullSim, FastSim). All jets need to fulfill the requirement that no neighboring jet is closer than $\Delta R = 0.5$ avoiding overlapping jets. Otherwise, no additional selection is applied. Jets have transverse momenta as low as $\pt=15\GeV$ and the range of pseudorapidity is fully inclusive. The final data sample used for training consists of roughly 6 million jet triplets.

\section{Method}
\label{sec:method}
A vector of analysis observables simulated by the FastSim chain is defined as \xfast and can be compared to \xfull, the corresponding FullSim output. A fully-connected feed-forward neural network is trained to provide an output \xrefined for a given input vector \xfast. The aim is to establish a \emph{refined} version of the FastSim data sample, which is more similar to the FullSim output, i.e., more accurate.

The analysis observables used in this study are four jet flavor tagging observables available in the CMS NanoAOD data analysis format:
\begin{equation}
    \vec{x} = \begin{pmatrix} \B & \CvB & \CvL & \QG \end{pmatrix}^\intercal\,.
\end{equation}
They are calculated from the output of the DeepJet algorithm~\cite{Bols_2020}, a multiclass neural network with six output nodes, activated with a softmax function. The nodes correspond to jets containing hadronically (leptonically) decaying b hadrons (labeled b (lepb)), jets containing two b hadrons (bb), and jets from c quarks (c), light quarks (uds), and gluons (g). From those values, the four discriminator observables are calculated as
\begin{equation}
    \begin{aligned}
        \B = \bnode + \bbnode + \lepbnode\,,\;
        \CvB = \frac{\cnode}{\cnode + \bnode + \bbnode + \lepbnode}\,,\;
        \CvL = \frac{\cnode}{\cnode + \udsnode + \gnode}\,,\;
        \QG = \frac{\gnode}{\gnode + \udsnode}\,.
    \end{aligned}
    \label{eq:deepjet}
\end{equation}
The refinement network is parametrized by a vector of parameters $\vec{y}$, including the true transverse momentum of the jet $\pt^{\text{GEN}}$, the true pseudorapidity $\eta^{\text{GEN}}$, and the true hadron flavor.

\subsection{Network architecture}
\label{sec:network architecture}
The neural network is composed of five residual blocks, which each consist of two linear layers with 1024 nodes and a skip connection adding the input of the first layer back to the output of the second layer. This ResNet-like architecture~\cite{7780459} is motivated by the fact that the FastSim output is already a good approximation of the FullSim output and we only want to apply a residual correction. The Leaky ReLU activation function (with a slope of 0.01 for negative input values) is used between layers whereas no activation function is applied to the output of the very last layer. During training, dropout with a rate of 50\% is used to avoid overtraining.

Additionally, the input variables and parameters are transformed before they are passed to the network. The logit transformation is used for the DeepJet discriminators. For the true transverse momentum $\pt^{\text{GEN}}$, values are first scaled to the interval $(0, 1)$ by applying the transformation {$\tanh{\frac{\pt^{\text{GEN}}}{200}}$} and then also logit-transformed. The true hadron flavor is one-hot-encoded and the true pseudorapidity $\eta^{\text{GEN}}$ is not transformed.

\begin{figure}[htb!]
    \centering
    \includegraphics[width=0.8\textwidth]{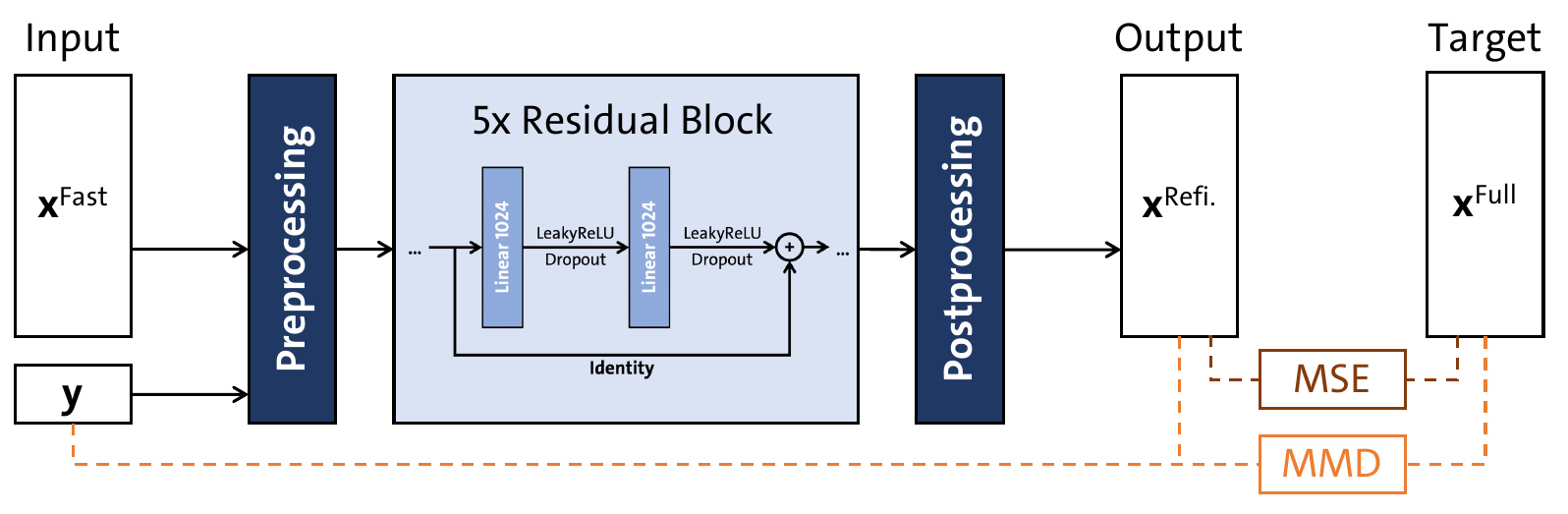}
    \caption{The refinement network is built from residual blocks embedded in pre- and postprocessing layers. The inputs to the loss functions are indicated by the dashed lines.}
    \label{fig:nnsketch}
\end{figure}

A postprocessing layer is appended to the network to apply the inverse transformations and to enforce the constraint that the refined DeepJet discriminators (B, CvB, CvL, QG) have to be constructed such that the sum of the original DeepJet output nodes (b, bb, lepb, c, uds, g) is equal to unity. This is done by analytically calculating the values of the output nodes according to Eq.~\eqref{eq:deepjet} and then dividing each value by the sum. The network is implemented using the PyTorch package~\cite{NEURIPS2019_9015} and its architecture is summarized in Fig.~\ref{fig:nnsketch}.

\subsection{Loss terms}
\label{sec:loss terms}
The most widely used loss function for regression tasks is the mean-squared-error (L2) loss, defined for a batch of $n$ output-target pairs as:
\begin{equation}
    \mse = \frac{1}{n} \sum_{i=1}^{n} \left( \xrefined_i-\xfull_i \right)^2\,.
    \label{eq:mse}
\end{equation}
To devalue outliers while keeping a smooth behavior around 0, we use the Huber loss function, which is a combination of the mean-squared-error and mean-absolute-error:

\begin{gather}
    \huber = \frac{1}{n} \sum_{i=1}^{n}h_i\,,\\
    h_i = \left\{\begin{matrix}
        0.5 \left( \xrefined_i-\xfull_i \right)^{2} & \text{if} \; \left| \xrefined_i-\xfull_i  \right| < \delta \\
        \delta \left( \left| \xrefined_i-\xfull_i  \right| - 0.5 \delta \right) & \text{otherwise}
        \end{matrix}\right. \qquad \text{with} \; \delta = 0.1 \,.
    \label{eq:huber}
\end{gather}
Both of these functions measure distances between fixed output-target pairs and thus serve to render each FastSim jet more similar to the corresponding FullSim jet. While this addresses deterministic biases, the independent stochasticity in both simulation chains prevents the refinement of full distributions. Training only with such a loss term consistently leads to a regression to the mean as fluctuations towards small (large) values are systematically corrected to larger (smaller) values. This problem is overcome by using an ensemble-based loss function, the maximum mean discrepancy (MMD)~\cite{JMLR:v13:gretton12a}.

The sample estimate of the MMD is calculated by evaluating a kernel function for all possible pairs of output-output, target-target, and output-target samples defined by $\vec{x}$. The parameter vector $\vec{y}$ is concatenated to the vectors $\vec{x}$ to include correlations between the variables and parameters:
\begin{equation}
    \begin{aligned}
        \mmd_\sigma = & \frac{1}{n^2} \sum_{i=1}^{n}  \sum_{j=1}^{n} k_\sigma \left( \begin{pmatrix}\xrefined_i \\ \vec{y}_i \end{pmatrix}, \begin{pmatrix}\xrefined_j \\ \vec{y}_j \end{pmatrix} \right)
        + \frac{1}{n^2} \sum_{i=1}^{n}  \sum_{j=1}^{n} k_\sigma \left( \begin{pmatrix}\xfull_i \\ \vec{y}_i \end{pmatrix}, \begin{pmatrix}\xfull_j \\ \vec{y}_j \end{pmatrix} \right) \\
        & - \frac{2}{n^2} \sum_{i=1}^{n}  \sum_{j=1}^{n} k_\sigma \left( \begin{pmatrix}\xrefined_i \\ \vec{y}_i \end{pmatrix}, \begin{pmatrix}\xfull_j \\ \vec{y}_j \end{pmatrix} \right)\,.
    \end{aligned}
    \label{eq:mmd}
\end{equation}
We use a Gaussian kernel function $k_\sigma(x_1,x_2)= \exp{\left( -\frac{1}{\sigma} \lVert x_1 - x_2 \rVert^{2} \right)}$ with bandwidth $\sigma$. The values for $\mmd_\sigma$ with five different bandwidths are added to make up the MMD loss: 
\begin{equation}
    \mmd = \sum\nolimits_{\sigma\,\in\,\{\frac{\sigma_0}{100},\, \frac{\sigma_0}{10},\, \sigma_0,\,  10\sigma_0,\, 100\sigma_0\}} \mmd_\sigma\,,
\end{equation}
where the central bandwidth is the mean L2 distance between all considered pairs:
\begin{equation}
    \begin{aligned}
        \sigma_0 = & \frac{1}{4} \left( \frac{1}{n^2} \sum_{i=1}^{n}  \sum_{j=1}^{n} \left\lVert \begin{pmatrix}\xrefined_i \\ \vec{y}_i \end{pmatrix} - \begin{pmatrix}\xrefined_j \\ \vec{y}_j \end{pmatrix} \right\rVert^{2} \right.
        \left. + \frac{1}{n^2} \sum_{i=1}^{n}  \sum_{j=1}^{n} \left\lVert \begin{pmatrix}\xfull_i \\ \vec{y}_i \end{pmatrix} - \begin{pmatrix}\xfull_j \\ \vec{y}_j \end{pmatrix} \right\rVert^{2} \right. \\
        & \left.  + \frac{2}{n^2} \sum_{i=1}^{n}  \sum_{j=1}^{n} \left\lVert \begin{pmatrix}\xrefined_i \\ \vec{y}_i \end{pmatrix} - \begin{pmatrix}\xfull_j \\ \vec{y}_j \end{pmatrix} \right\rVert^{2} 
        \right)\,.
    \end{aligned}
\end{equation}

\subsection{MDMM algorithm}
\label{sec:mdmm algorithm}
The above two losses are partially correlated, but cannot, in general, be extremized simultaneously. To arrive at an optimal set of influences from each loss, we use an algorithm called the Modified Differential Method of Multipliers (MDMM), introduced in Ref.~\cite{NIPS1987_a87ff679}. The MDMM reformulates the training as a Lagrangian optimization process. Identifying a primary loss $f(\theta)$ and an additional loss $g(\theta)$, which both depend on the network parameters $\theta$, the Lagrangian becomes
\begin{equation}
    \mathcal{L}(\theta, \lambda)=f(\theta) - \lambda \left( \varepsilon - g(\theta) \right)\,.
\end{equation}
Hence, the objective is to minimize the primary loss subject to the constraint $g(\theta) = \varepsilon$. The algorithm uses stochastic gradient descent for the network parameters $\theta$ but gradient ascent for the Lagrange multiplier $\lambda$, meaning that the weight of the additional loss is dynamically updated during the training. Additionally, a damping term $0.5\left( \varepsilon - g(\theta) \right)^2$ is added to the Lagrangian to ensure convergence. The implementation of the MDMM algorithm is adapted from the mdmm package~\cite{mdmmSW}.

\subsection{Training}
\label{sec:training}
\begin{figure}[htb!]
    \centering
    \includegraphics[width=0.5\textwidth]{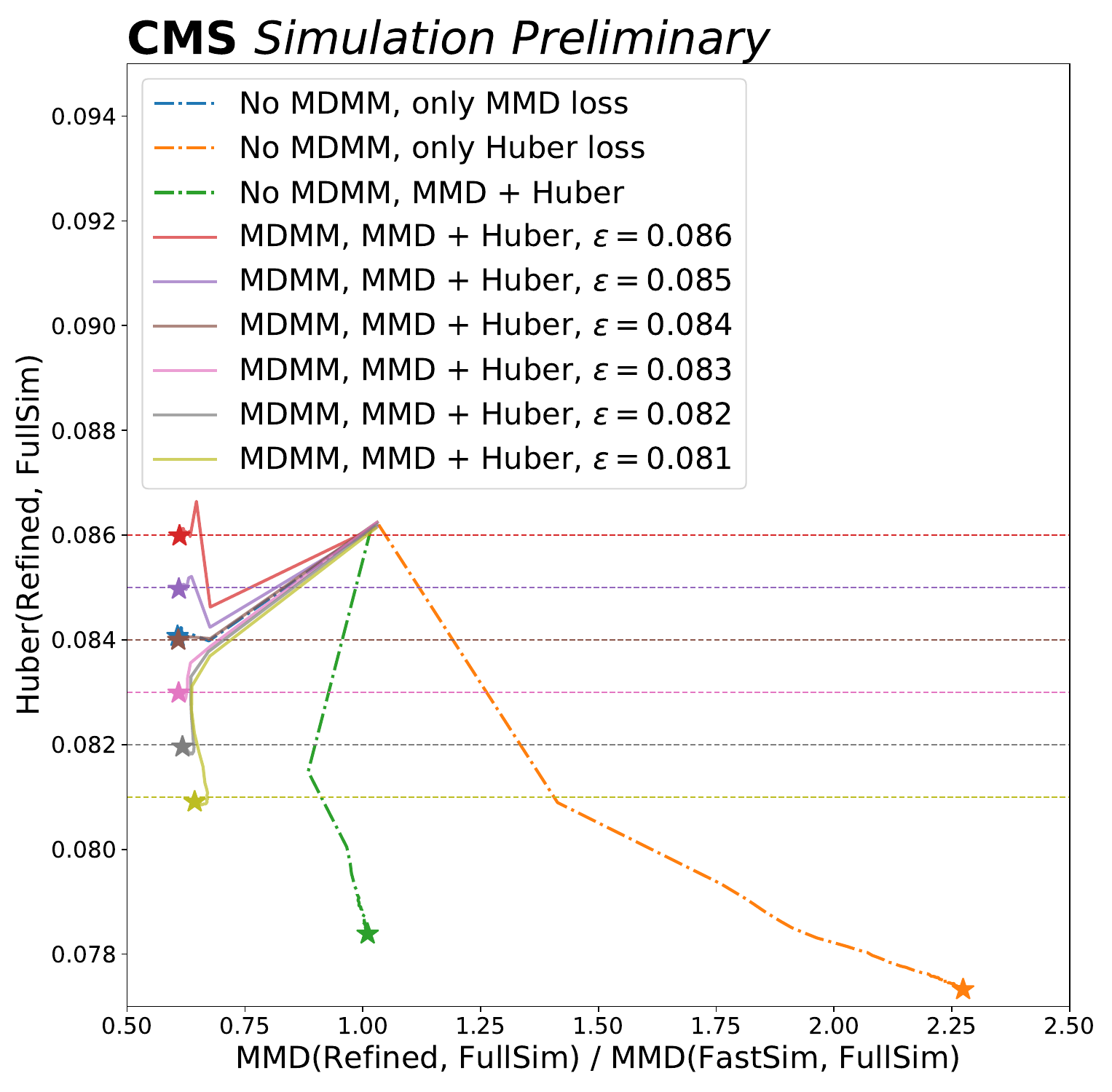}
    \caption{Convergence of different network trainings with and without the MDMM algorithm shown in the plane of the two loss terms, Huber and MMD. The training endpoints are marked with stars and for trainings using the MDMM algorithm, the values of $\varepsilon$ are indicated by horizontal lines.}
    \label{fig:pareto}
\end{figure}

The network is trained with 500 batches of 4096 jet triplets for 100 epochs. An additional 1000 batches of 4096 jet triplets are used as validation and test datasets with a 50/50 split. Different loss schemes are investigated both with and without the MDMM algorithm: training versions without MDMM are performed with only the MMD loss, only the Huber loss, and with a simple sum of the two. For the training versions with MDMM, MMD is taken as the primary loss while Huber is the additional loss, and different values of $\varepsilon$ are explored. The MDMM algorithm uses the Adam optimizer with a learning rate of 0.0001. Fig.~\ref{fig:pareto} shows the convergence of all training versions in the plane of the two loss functions. The value of MMD(Refined, FullSim) is normalized to the baseline value of MMD(FastSim, FullSim), meaning that the starting points are all very close to 1 and values smaller than 1 indicate improved agreement between the distributions.

The version with only the Huber loss, while improving the agreement between the fixed corresponding jet-jet pairs (the value on the y-axis), drastically worsens the agreement between the ensembles (given by the value on the x-axis) demonstrating regression to the mean as expected. Also, a naive addition of MMD and Huber does not lead to a better distribution-level agreement. If, however, the network is trained using only the MMD loss, an improvement on both axes can be observed. Yet, the point on the Pareto front (the set of all optimal solutions) to which the training converges cannot be chosen directly in such a setup. This could be achieved indirectly by experimenting with different fixed weights in the sum of the two loss terms, but when using the MDMM algorithm, the Pareto front can be scanned directly by choosing different values for $\varepsilon$. We observe a convex Pareto front indicating the tradeoff between the two loss terms. In the following section, the results are shown for the training with MDMM using \mbox{$\varepsilon = 0.084$}, which converges close to the MMD-only training.

\section{Results}
\label{sec:results}
\begin{figure}[htb!]
    \centering
        \includegraphics[width=0.35\textwidth]{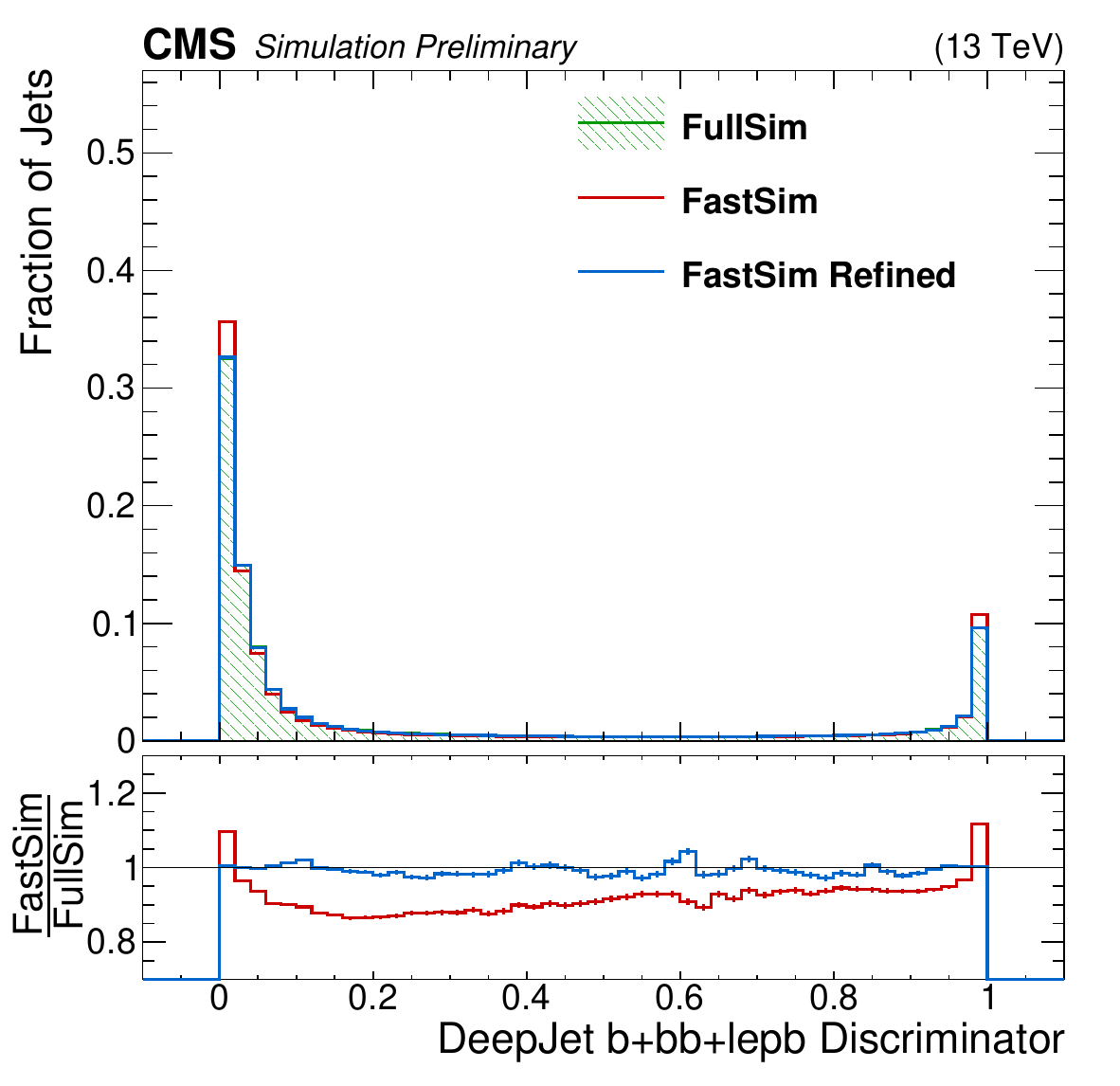}
        \includegraphics[width=0.35\textwidth]{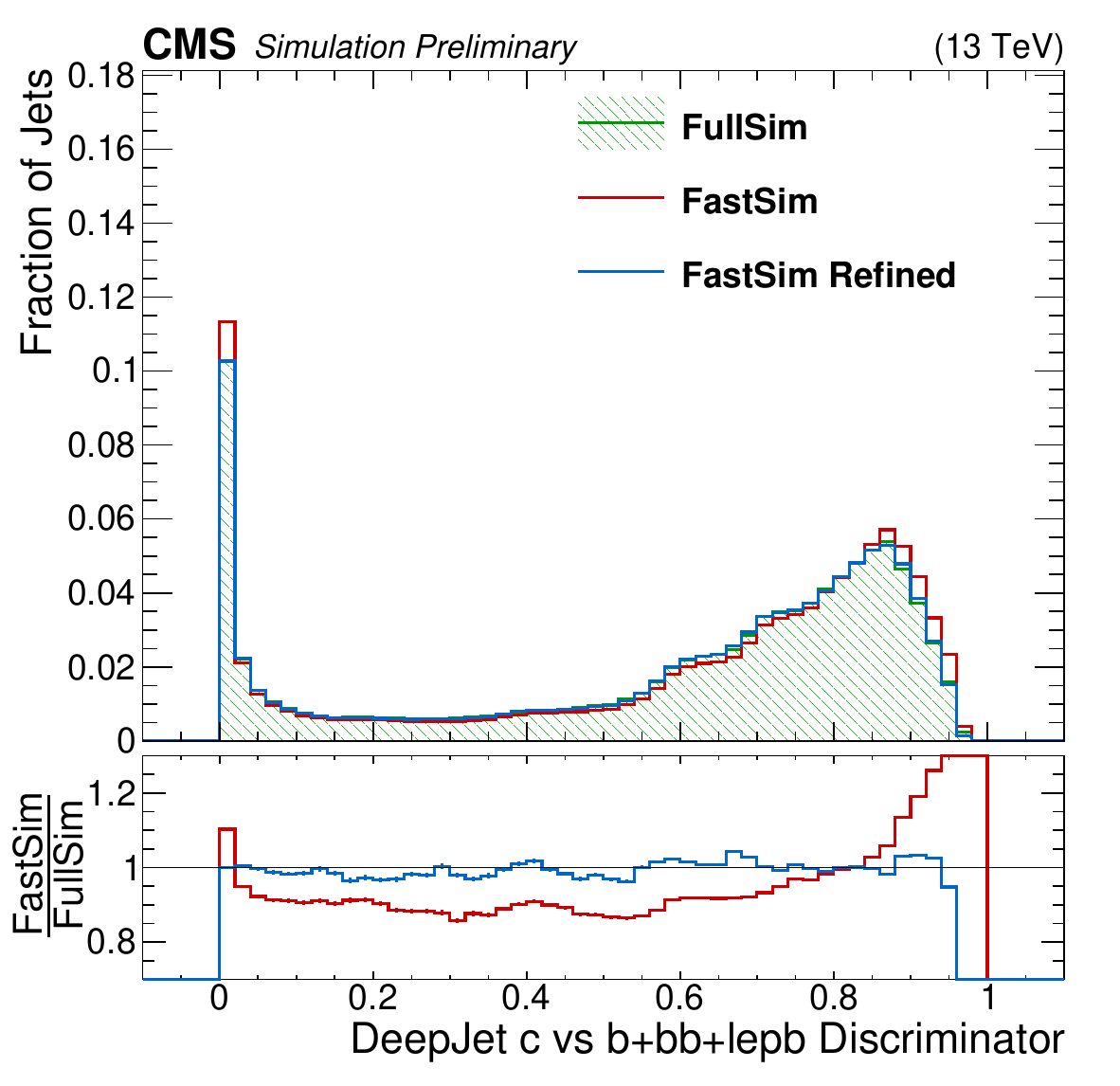}\\
        \includegraphics[width=0.35\textwidth]{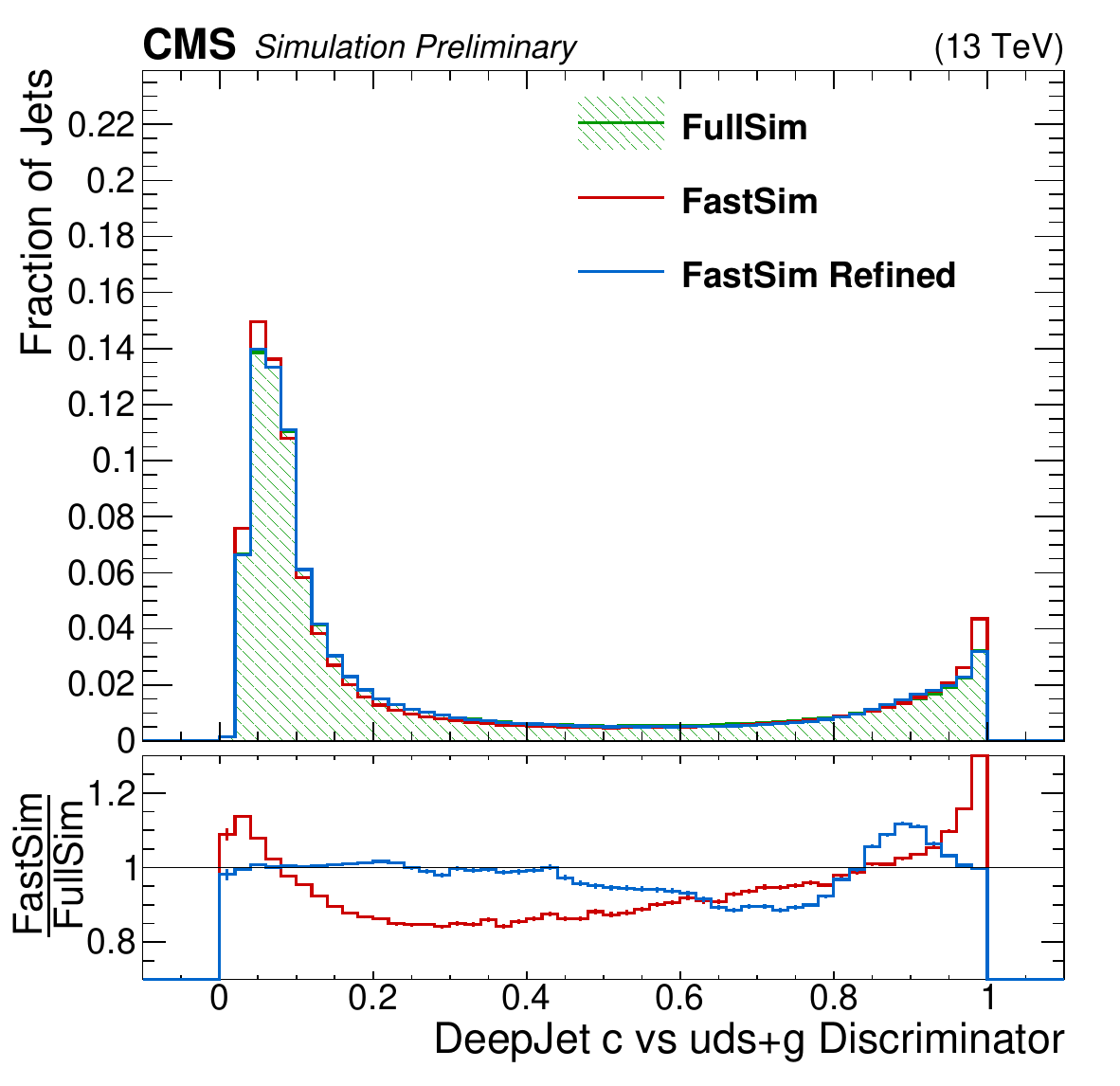}
        \includegraphics[width=0.35\textwidth]{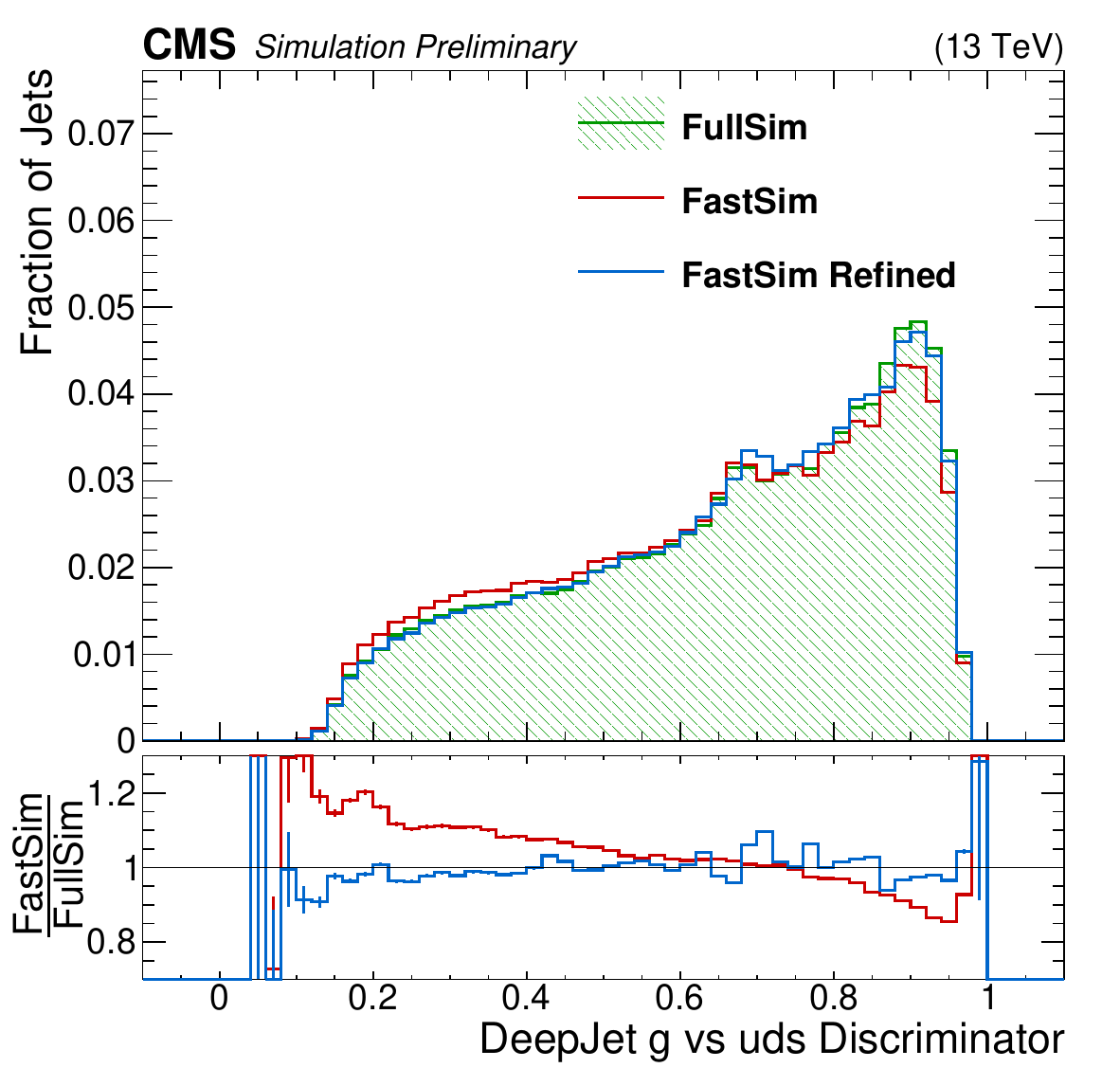}
    \caption{The distributions of the four DeepJet discriminators B (upper left), CvB (upper right), CvL (lower left), and QG (lower right) for FullSim, FastSim, and the refined version of FastSim.}
    \label{fig:results}
\end{figure}

\begin{figure}[htb!]
    \centering
    \includegraphics[width=\textwidth]{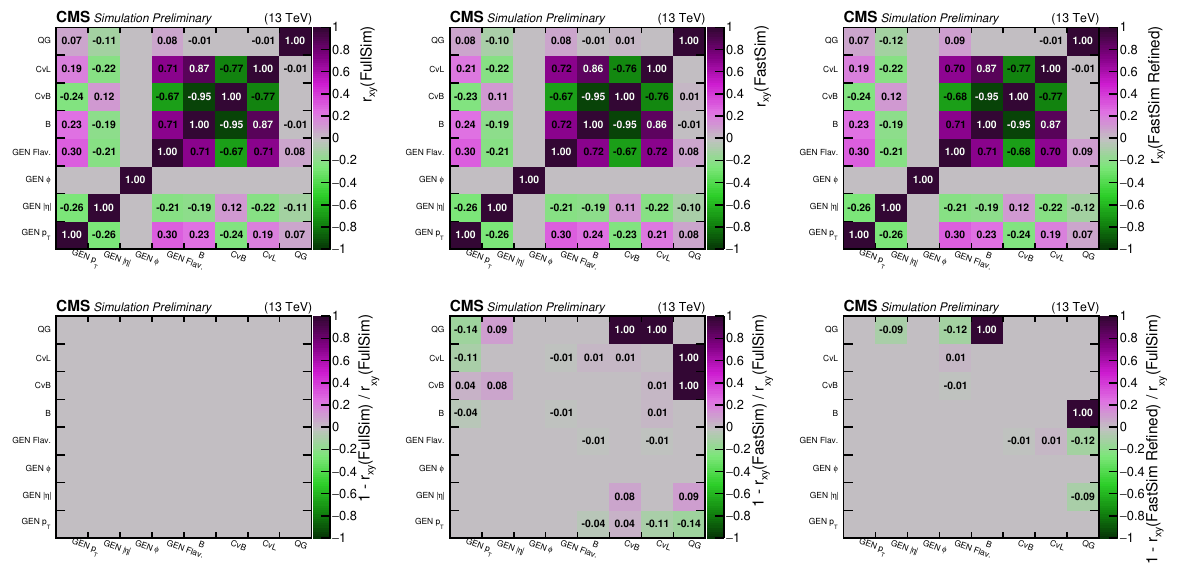}
    \caption{Upper: the Pearson correlation coefficients (rounded to two digits) for all pairs of variables and parameters for FullSim (left), FastSim (center), and refined FastSim (right). Lower: the relative difference between the correlation factors for each simulation and the respective FullSim value with the same axes. All entries in the bottom left plot are zero by construction and a relative difference of 100\% indicates cases where the correlation factor for FullSim or for (refined) FastSim in the upper plot is zero.}
    \label{fig:correlation factors}
\end{figure}

Fig.~\ref{fig:results} shows the distributions of the four DeepJet discriminators for the three datasets: FullSim, FastSim, and refined FastSim. It can be seen that the refinement improves the accuracy of FastSim, as quantified by the agreement with the FullSim output. Furthermore, Fig.~\ref{fig:correlation factors} shows that correlations within the considered variables and between the variables and parameters have better agreement after refinement.

We also compute values for the Fr\'{e}chet Physics Distance (FPD) and Kernel Physics Distance (KPD) introduced in Ref.~\cite{Kansal:2022spb} and calculated with the JetNet package~\cite{Kansal:2021cqp,raghav_kansal_2023_7778868}, shown in Table~\ref{tab:metrics}. The metrics are computed in the four-dimensional space of the original (not logit-transformed) DeepJet discriminators. The metric values for refined FastSim are similar in magnitude to a baseline comparison of FullSim to itself, an order of magnitude improvement compared to unrefined FastSim. The trained network is also evaluated on a sample of 6 million jet triplets from simulated top quark pair production (\ttbar) events and the corresponding metric values are stated in Table~\ref{tab:metrics}. This indicates that the method generalizes to different event topologies.

\begin{table}[htb!]
    \centering
    \topcaption{The Fr\'{e}chet and Kernel Physics Distance metrics computed using the DeepJet discriminators to quantify the agreement between the different simulation approaches for multiple physics processes. The last row serves as a baseline, computed by comparing the first half of the FullSim sample to the second half.}
    \cmsTable{
    \begin{tabular}{l B B B B}
        FullSim vs. & \multicolumn{4}{c}{FPD $\times 10^3$} & \multicolumn{4}{c}{KPD $\times 10^3$} & \multicolumn{4}{c}{FPD $\times 10^3$ (\ttbar)} & \multicolumn{4}{c}{KPD $\times 10^3$ (\ttbar)} \\
        \hline
        FastSim & 0&.801 & 0&.046 & 1&.07 & 0&.58  & 0&.540 & 0&.036 & 0&.927 & 0&.448 \\
        FastSim Refined & 0&.071 & 0&.025 & 0&.083 & 0&.418 & 0&.065 & 0&.025 & -0&.127 & 0&.164  \\
        FullSim & 0&.061 & 0&.029 & -0&.024 & 0&.250 & 0&.061 & 0&.024 & -0&.119 & 0&.167 
    \end{tabular}
    }
    \label{tab:metrics}
\end{table}

\section{Conclusions}
\label{sec:conclusions}
We have introduced a novel approach to refine the output of the CMS FastSim chain to better match the FullSim chain. A regression neural network is trained with the maximum mean discrepancy (MMD) as the primary loss, formulating an ensemble-based training objective not relying on fixed output-target pairs. Additionally, we incorporate the pair-based Huber loss as a constraint on the training via the modified differential method of multipliers. We apply the method to a sample of jets simulated with both FastSim and FullSim, attempting the refinement of four DeepJet flavor-tagging discriminators. The results show a clear improvement in agreement with the FullSim output in one-dimensional projections, linear correlation coefficients, and dedicated metrics. Besides the improved accuracy, another advantage of this method is the absence of weights, which would reduce the statistical power of the simulated sample. Stability and convergence were observed for all network trainings. The method can be straightforwardly extended to other variables, and a refinement of simulation directly to data is possible using the MMD loss, which does not require labeled pairs of events or objects.

\bibliography{main}

\end{document}